\documentclass[useAMS,usenatbib]{mn2e}

\pdfoutput=1

\usepackage{epsf, epsfig}

\usepackage[shortcuts,acronyms]{glossaries-extra}		
\usepackage{enumitem}
\usepackage{amssymb}
\usepackage{siunitx}									

\def\href#1#2{#2}
\def\urlinner#1{#1\endgroup}
\def\url{\begingroup\def\do##1{\catcode`##1 12 }%
  \do\\\do\$\do\&\do\#\do\^\do\_\do\%\do\~ \ttfamily \urlinner}

\newabbr{DM}{DM}{dark matter}
\newabbr{CDM}{CDM}{Cold Dark Matter}
\newabbr{GC}{GC}{globular cluster}
\newabbr{UCD}{UCD}{ultra-compact dwarf}
\newabbr{COM}{COM}{centre of mass}
\newabbr{MW}{MW}{Milky Way}

\expandafter\def\csname rundmg105 \endcsname{O6-105}
\expandafter\def\csname rundmg106 \endcsname{O1-106}
\expandafter\def\csname rundmg108 \endcsname{O7-108}
\expandafter\def\csname rundmg109 \endcsname{O2-109}
\expandafter\def\csname rundmg110 \endcsname{O8-110}
\expandafter\def\csname rundmg114 \endcsname{O4-114}
\expandafter\def\csname rundmg115 \endcsname{O5-115}
\expandafter\def\csname rundmg116 \endcsname{O3-116}
\expandafter\def\csname rundmg13 \endcsname{O9-13}
\expandafter\def\csname rundmg14 \endcsname{O10-14}
\expandafter\def\csname rundmg118 \endcsname{S1}
\expandafter\def\csname rundmg118a \endcsname{S1}
\expandafter\def\csname rundmg119 \endcsname{S2}
\expandafter\def\csname rundmg120 \endcsname{S3}
\expandafter\def\csname rundmg121 \endcsname{S4}
\expandafter\def\csname rundmg122 \endcsname{S5}
\expandafter\def\csname rundmg123 \endcsname{S6}
\expandafter\def\csname rundmg124 \endcsname{S7}
\expandafter\def\csname rundmg125 \endcsname{L8}
\expandafter\def\csname rundmg126 \endcsname{S9}
\expandafter\def\csname rundmg127 \endcsname{L10}
\expandafter\def\csname rundmg128 \endcsname{S11\_O}
\expandafter\def\csname rundmg129 \endcsname{S12}
\expandafter\def\csname rundmg130 \endcsname{S13\_O}
\expandafter\def\csname rundmg131 \endcsname{S14\_O}
\expandafter\def\csname rundmg133 \endcsname{S15\_O}
\expandafter\def\csname rundmg134 \endcsname{S16\_O}
\expandafter\def\csname rundmg135 \endcsname{S17}
\expandafter\def\csname rundmg136 \endcsname{S18}

\title[Formation of massive globular clusters]{Formation of massive globular clusters with dark matter and its implication on dark matter annihilation} 
\author[H. Wirth, K. Bekki and K. Hayashi]
{Henriette Wirth${}^1$\thanks{E-mail: henri-ette\_w@web.de},
Kenji Bekki${}^2$ and
Kohei Hayashi${}^3$ \\ 
${}^1$Elektronische Fahrwerksysteme GmbH,
Dr.-Ludwig-Kraus-Str. 6,
85080 Gaimersheim, Germany
\\
${}^2$ICRAR, M468,
The University of Western Australia,
35 Stirling Hwy, Crawley,
Western Australia 6009, Australia
\\
${}^3$Institute for Cosmic Ray Research,
The University of Tokyo,
Chiba 277-8583, Japan}

\begin{document}

\date{Accepted, Received 2005 February 20; in original form }

\pagerange{\pageref{firstpage}--\pageref{lastpage}} \pubyear{2005}

\maketitle

\label{firstpage}

\begin{abstract}
Recent observational studies of $\gamma$-ray emission from massive \acp{GC} have revealed possible evidence of \ac{DM} annihilation within \acp{GC}.
It is, however, still controversial whether the emission comes from \ac{DM} or from milli-second pulsars.
We here present the new results of numerical simulations, which demonstrate that \acp{GC} with \ac{DM} can originate from nucleated dwarfs orbiting the ancient \ac{MW}.
The simulated stripped nuclei (i.e., \acp{GC}) have the central \ac{DM} densities 
ranging  from 0.1 to several ${\rm M_\odot pc^{-3}}$, depending on the orbits and the masses of the host dwarf
galaxies.  However,  \acp{GC} born outside the central regions of their hosts can have no/little \ac{DM} after
their hosts are destroyed and the \acp{GC} become the Galactic halo \acp{GC}. These results
suggest that only \acp{GC} originating from stellar nuclei of dwarfs can possibly have \ac{DM}.
We further calculate the expected $\gamma$-ray emission from these simulated \acp{GC} and compare them to observations of $\omega$ Cen.
Given the large range of \ac{DM} densities in the simulated \acp{GC}, we suggest that the recent
possible detection of \ac{DM} annihilation from \acp{GC} should be more carefully interpreted.
\end{abstract}

\begin{keywords}
globular clusters: general --
globular clusters: individual: Omega Centauri --
dark matter
\end{keywords}

\glsresetall

\begin{table*}
\centering
\begin{minipage}{180mm}
\centering
\caption{The physical properties of the simulated stripped nuclei (and GCs) at T=2.82 Gyr.
As described in section \ref{sec_Model_NucDwarf} we label our standard models with ``S'', our low-density models with ``L'' and we add an ``\_O'' if the model has an off-centre nucleus.}
\label{table_T20}
\begin{tabular}{r
                    *{4}{S[detect-weight,
                        mode=text,       
                        table-format=1.3]}
                    c
                    *{6}{S[detect-weight,
                        mode=text,       
                        table-format=1.3]}
                        }
    \hline
    &&&&&& {$R < 30~\rm pc$:}
    &&& {$R < 100~\rm pc$:}\vspace{2pt}\\
    ID 
    & {$M_{\rm nuc}$} 
    & {$R_{\rm nuc}$} 
    &  {$R_{ini}$}
    & {$R_{\rm peri}$}
    &  {$i$}
    & {$F_{\rm DM}$} 
    & {$F_{\rm s}$} 
    & {$\rho_{\rm DM}$}
    & {$F_{\rm DM}$} 
    & {$F_{\rm s}$} 
    & {$\rho_{\rm DM}$}
    \\
    & {($\times 10^{6} ~{\rm M}_{\odot}$)}
    &  {(pc)}
    &  {(kpc)}
    & {(kpc)}
    &  {($^{\circ}$)}
    & {($\times 10^{-3}$)} 
    & {($\times 10^{-3}$)} 
    & {($M_\odot {\rm pc}^{-3}$)}
    & {($\times 10^{-3}$)} 
    & {($\times 10^{-3}$)} 
    & {($M_\odot {\rm pc}^{-3}$)}
    \\\hline
       \csname rundmg118a \endcsname & 10.0 &  30.0 &  17.50&2.0 &  60 &    15.4 &     0.0 &   1.66 &    61.3 &     2.3 &   0.23\\
       \csname rundmg119 \endcsname & 30.0 &  50.0 &  35.00&8.5 &  60 &    17.3 &     0.1 &   5.08 &   118.9 &     5.2 &   1.34\\
       \csname rundmg120 \endcsname & 10.0 &  30.0 &  8.75&3.0 &  60 &    14.5 &     0.0 &   1.49 &    42.1 &     2.8 &   0.15\\
       \csname rundmg121 \endcsname &  1.0 &  30.0 &  17.50&7.0 &  60 &    13.6 &     4.0 &   0.19 &    67.3 &    40.5 &   0.03\\
       \csname rundmg122 \endcsname &  1.0 &  30.0 &  35.00&6.0 &  60 &    85.4 &    48.8 &   1.22 &  2888.2 &  1432.9 &   1.20\\
       \csname rundmg123 \endcsname & 10.0 &  30.0 &  17.50&2.5 &  60 &    15.4 &     0.0 &   1.66 &    61.3 &     2.3 &   0.23\\
       \csname rundmg124 \endcsname & 10.0 &  30.0 &  17.50&2.0 &  30 &    13.8 &     0.0 &   1.33 &    43.8 &     2.3 &   0.14\\
       \csname rundmg125 \endcsname & 10.0 &  30.0 &  17.50&2.0 &  60 &     4.5 &     0.0 &   0.47 &    21.5 &     4.8 &   0.08\\
       \csname rundmg126 \endcsname & 10.0 &  30.0 &  8.75&2.0 &  30 &    13.0 &     0.6 &   1.33 &    81.4 &    13.2 &   0.27\\
       \csname rundmg127 \endcsname & 10.0 &  30.0 &  8.75&1.0 &  60 &     4.5 &     0.0 &   0.45 &    14.0 &     1.0 &   0.05\\
       \csname rundmg128 \endcsname &  1.0 &  30.0 &  17.50&7.0 &  60 &     0.5 &     6.9 &   0.01 &    45.9 &    89.6 &   0.02\\
       \csname rundmg129 \endcsname & 10.0 &  30.0 &  5.25&5.0 &  30 &    19.8 &     1.4 &   1.05 &   308.5 &    52.2 &   0.47\\
       \csname rundmg130 \endcsname &  1.0 &  30.0 &  17.50&7.0 &  60 &    11.7 &     5.4 &   0.16 &    33.4 &    53.7 &   0.01\\
       \csname rundmg131 \endcsname & 30.0 &  50.0 &  17.50&2.0 &  60 &     0.6 &     0.0 &   0.17 &     5.3 &     0.6 &   0.06\\
       \csname rundmg133 \endcsname & 10.0 &  30.0 &  17.50&0.5 &  60 &     0.0 &     0.0 &   0.00 &     3.6 &     1.8 &   0.01\\
       \csname rundmg134 \endcsname & 30.0 &  50.0 &  35.00&9.0 &  60 &     0.0 &     0.2 &   0.01 &    30.6 &     8.6 &   0.33\\
       \csname rundmg135 \endcsname &  0.1 &  10.0 &  17.50&7.0 &  60 &   121.4 &  2756.9 &   0.18 &  4058.0 & 45882.6 &   0.17\\
       \csname rundmg136 \endcsname & 10.0 &  30.0 &  17.50&3.0 &  60 &    12.8 &     0.0 &   1.35 &    56.0 &     1.5 &   0.21\\
    \hline
    \end{tabular}

\end{minipage}
\end{table*}

\section{Introduction}

The Galactic \ac{GC} $\omega$ Cen has a number of very unique characteristics,
such as the very large mass \citep[e.g.][]{1995A&A...303..761M}, retrograde orbits \citep[e.g.][]{1999AJ....117.1792D},
and multiple distinct subpopulations \citep[e.g.][]{2018ApJ...853...86B}. These unique properties have been
investigated both observationally and theoretically \citep[e.g.][]{1986A&A...166..122M, 2013MNRAS.436.2598W, 2019MNRAS.482.5138B}.
Recent observations detected $\gamma$-ray emission from massive \acp{GC} like $\omega$ Cen and 47 Tucanae \citep{2010A&A...524A..75A, 2020ApJ...888L..18D}.
The source of this $\gamma$-ray emission is still subject to debates.
The most popular hypotheses include the presence of millisecond pulsars \citep{2010A&A...524A..75A, 2020ApJ...888L..18D} and \ac{DM} annihilation \citep{2016ConPh..57..496G, 2019arXiv190708564B}.
For a direct comparison of the two possibilities see \cite{2019arXiv190706682R}.
If the source of the $\gamma$-rays is \ac{DM}, then the question would be: Where did the \ac{DM} come from?

It has been suggested that nucleated dwarf galaxies can be transformed into 
massive \acp{GC} like $\omega$ Cen and \acp{UCD} due to tidal stripping of the dwarfs by the strong gravitation field of their host environments \citep[``galaxy threshing'';][from here on BF03]{2001ApJ...552L.105B,2003MNRAS.346L..11B}.
In particular $\omega$ Cen has been proposed to be the tidally stripped nucleus of a dwarf galaxy (BF03).
This is further supported by its density profile \citep{2004ApJ...616L.107I}.
However, previous simulations did not include a \ac{DM} halo \citep{2004ApJ...616L.107I}, which is known to exist in dwarf galaxies \citep{2004IAUS..220..377K, Kormendy_2016, 2019arXiv191205352D}.
Therefore, if $\omega$ Cen is the nucleus of a tidally stripped dwarf, could there be any \ac{DM} be left over from the progenitor galaxy?
Similar suggestions have been made to explain the elevated mass to luminousity ratio in \acp{UCD} \citep{2011MNRAS.412.1627C}.
Dark stellar clusters around Centaurus A are also believed to be remnants of a stripped dwarf \citep{2016ApJ...832...88B}.

The purpose of this paper is to investigate how much \ac{DM} can be left in the stripped stellar galactic nuclei that can be progenitors of massive GCs.
To this end we run a set of simulations of dwarfs with different initial parameters on different orbits around the \ac{MW}.
We also calculate the J-Factor resulting from our final \ac{DM} distribution and 
discuss whether or not the observed flux of gamma ray emission in $\omega$ Cen can be really explained by annihilation of \ac{DM} gravitationally trapped by the GC. 

\section{The model}

\subsection{Nucleated dwarfs orbiting the Galaxy}
\label{sec_Model_NucDwarf}

The present code for direct Nbody simulations of nucleated dwarf
galaxies is essentially the same as the one used in \citep[BT16]{2016ApJ...831...70B} in which
the dynamical evolution of GCs in a dwarf galaxy
orbiting the Galaxy is investigated.
Since the details of the simulation code are given in BT16,
we here briefly describe the code.
It should be stressed here that the adopted code 
does not allow us to investigate how gas and star formation
can control the evolution of interacting dwarfs and stellar nuclei
which were investigated in our other works
\citep[e.g.][]{2007PASA...24...77B,2007PASA...24...21B,2019A&C....2800286B}.
The \ac{DM} halo with the total mass of $M_{\rm h}$ in a nucleated dwarf galaxy
 is represented by
the `NFW' one \citep{1996ApJ...462..563N}
with a central cusp predicted by the \ac{CDM}  model:
\begin{equation}
{\rho}(r)=\frac{\rho_{0}}{(r/r_{\rm s})(1+r/r_{\rm s})^2},
\end{equation}
where $r$,  $\rho_{0}$,  and $r_{\rm s}$ are the distance from the center
of the cluster, the central density, and the scale-length of the dark halo,
respectively.
The virial radius ($r_{\rm vir}$),  the scale radius ($r_{\rm s}$),
and the `$c$' parameter (=$r_{\rm vir}/r_{\rm s}$)
are chosen such that the values
are consistent with recent cosmological simulations
for the adopted $M_{\rm h}$
\citep{2007MNRAS.381.1450N}.

In order to estimate the total mass and density of \ac{DM}
in the stripped stellar nuclei ($R<100$ pc) in a much better way,
 we here adopt the  following original setup for the dwarf's \ac{DM} halo:
We first divide the
\ac{DM} halo into two regions with $R \ge 100$ pc (``outer'')
and $R<100$ pc (``inner'') and use a particle mass ($m_{\rm dm}$) of only $0.02$ times that of the outer particles
and a factor 50 shorter time step width ($\Delta t$) for the inner halo.
The details of this new method will be discussed extensively in Bekki et al. (2020).
Here, we investigate models (labelled as \csname rundmg118a \endcsname\ etc)
in which $m_{\rm dm}=200~{\rm M}_{\odot}$ and $\Delta t = {10^4~\rm yr}$
were adopted for the inner halo, so that
we can resolve the inner 10pc-scale dynamical evolution of
nucleated dwarf galaxies.

The nucleated dwarf is assumed to be as a  bulge-less disk galaxy
with the total stellar mass of $M_{\rm s}$ and the size of $R_{\rm s}$.
The radial ($R$) and vertical ($Z$) density profiles of the stellar disk are
assumed to be proportional to $\exp (-R/R_{0}) $ with scale
length $R_{0} = 0.2R_{\rm s}$ and to ${\rm sech}^2 (Z/Z_{0})$ with scale
length $Z_{0} = 0.04R_{\rm s}$ , respectively.
The initial radial and azimuthal
velocity dispersions are assigned to the disc component according to
the epicyclic theory with Toomre's parameter $Q$ = 1.5. 
The stellar disk is assumed to have a stellar nucleus with a mass 
of $M_{\rm nuc}$ and a $5 \times$ scale radius of $R_{\rm nuc}$. The nucleus is represented
by a Plummer model with the free parameters $M_{\rm nuc}$ and $R_{\rm nuc}$.
Our dwarf galaxy
models have $M_{\rm dm}=10^{10} {\rm M}_{\odot}$,
$M_{\rm s}=1.2 \times 10^8 {\rm M}_{\odot}$, $R_{\rm s}={1.3~\rm kpc}$ and mostly $M_{\rm nuc}=10^7 {\rm M}_{\odot}$ and $R_{\rm nuc}={30~\rm pc}$, which is reasonable
for the formation of massive \acp{GC} from nucleated dwarfs \citep[e.g. BF03;][]{2012MNRAS.419.2063B}.
The mass resolution (and softening lengths) of the disk and stellar nucleus are ${1200}$ (${18.4}$) and ${1000 ~\rm M_\odot}$ (${0.3 ~\rm pc}$) respectively.

\subsection{The Milky Way model}

We investigated the ``young''
\ac{MW} models rather than the ``present-day''
ones (BT16) to discuss the formation of massive GCs from
stripped nuclei of dwarfs.
The Galaxy in the present \ac{MW} models is assumed to have a
{\it fixed} three-component gravitational potential
and the following  logarithmic \ac{DM} halo potential
is adopted for the Galaxy,
\begin{equation}
{\Phi}_{\rm halo}=v_{\rm halo}^2 \ln (r^2+d^2),
\end{equation}
where
$d$ = 12 kpc, $v_{\rm halo}$ = 93 km ${\rm s}^{-1}$ 
(instead of 131.5 km s$^{-1}$ suitable for the present-day Galaxy) and
$r$ is the distance from the center of the Galaxy.
The gravitational potential of the Galactic disk is represented by
a Miyamoto-Nagai potential \citep{1975PASJ...27..533M};
\begin{equation}
{\Phi}_{\rm disk}=-\frac{GM_{\rm disk}}{\sqrt{R^2 +{(a+\sqrt{z^2+b^2})}^2}},
\end{equation}
where $M_{\rm disk} = 1.0 \times 10^{10} M_{\odot}$,
(instead of $1.0 \times 10^{11} M_{\odot}$ for the present-day Galaxy),
and $a$ = 6.5 kpc, $b$ = 0.26 kpc,
and $R=\sqrt{x^2+y^2}$.
The following  spherical \cite{1990ApJ...356..359H} model is adopted for
the potential of the Galactic bulge;
\begin{equation}
{\Phi}_{\rm bulge}=-\frac{GM_{\rm bulge}}{r+c},
\end{equation}
where $M_{\rm bulge}$ =  3.4
$\times$ $10^{10}$ $M_{\odot}$,
and $c$ = 0.7 kpc.

To investigate how much \ac{DM} can be left in the stripped nucleus, we take the following steps:
First we evolve the dwarf through relaxation only (no tidal field) for 1.41 Gyr, then we expose the dwarf to the tidal field of the young \ac{MW}, where it is stripped.
We investigate dwarfs with different initial orbital velocities, dwarf positions, $M_{\rm nuc}$, and \ac{DM} properties of dwarf galaxies.
The orbits of the dwarfs with respect to the Galactic disk
have an inclination ($i$) of $30$ or $60$ degrees, and the stellar disks are inclined
by $45$ degrees with respect to the orbital planes. Although we 
mainly investigate ``standard'' models (labelled as ``S'')
with $R_{\rm vir}=17.9$ kpc and $c=16$ for \ac{DM},
we also investigate `low-density'  models (``L'')
with $R_{\rm vir}=17.9$ kpc and $c=8$.
Furthermore, we investigate model \csname rundmg128 \endcsname\ with a \ac{GC} a 200 pc outside of the dwarf's \ac{COM} and several models with a \ac{GC} 500 pc away from the \ac{COM}.
We will add an ``\_O'' to the label of those models.

\subsection{Estimation of $\gamma$-ray flux from \ac{DM} annihilation
in stripped nuclei}

Based on the mass and the density of \ac{DM} in a stripped nucleus,
we estimate the expected $\gamma$-ray emission stemmed from \ac{DM} annihilation using the CLUMPY code \citep{2019CoPhC.235..336H}.
To this end, we calculate the J-factor which is the integral of the squared \ac{DM} density along a line of sight over the cone with a solid angle $\Delta\Omega$:
\begin{equation}
J(\Delta\Omega)=\int_{\Delta\Omega}d\Omega\int_{l.o.s}d\ell \rho^2_{DM}(r(\ell,\Omega)),
\end{equation}
where $\ell$ is the line of sight coordinate.
Under the spherical symmetry assumption, we can rewrite $\Delta\Omega$ as $\Delta\Omega=2\pi\sin\theta d\theta$, where $\theta$ is the angular radius from the center of the object, and $r(\ell,\Omega)=\sqrt{\ell^2+d^2-2\ell d\cos\theta}$, where $d$ is the distance from the Sun ($d=5.4$~kpc for $\omega$ Cen).
Taking the value of $\rho_{DM}$ listed in the last column of Table \ref{table_T20}, we perform the integration over an angular radius $\Delta\Omega=0.7^{\circ}$.
For the estimation of the $\gamma$-ray energy spectrum, we adopt \ac{DM} particle mass $m_{DM}=31.4$~GeV estimated by \cite{2019arXiv190708564B} and the velocity-averaged annihilation cross-section $\log_{10}(<\sigma v>)={-27.3~[\rm cm^3 s^{-2}]}$, which is consistent with the upper limit on the cross section derived from a stacked analysis of dwarf spheroidal galaxies by Fermi-LAT data \citep[e.g.][]{2015PhRvL.115w1301A,2016MNRAS.461.2914H}.
We will use the following definitions: $F_{\rm DM} = \frac{M_{\rm DM}}{M_{\rm nuc}}$ and $F_{\rm s} = \frac{M_{\rm s}}{M_{\rm nuc}}$ and consider nuclei with $F_{\rm s}(R < 30~{\rm pc}) < 0.1$ to be stripped.


\begin{figure}
	\includegraphics{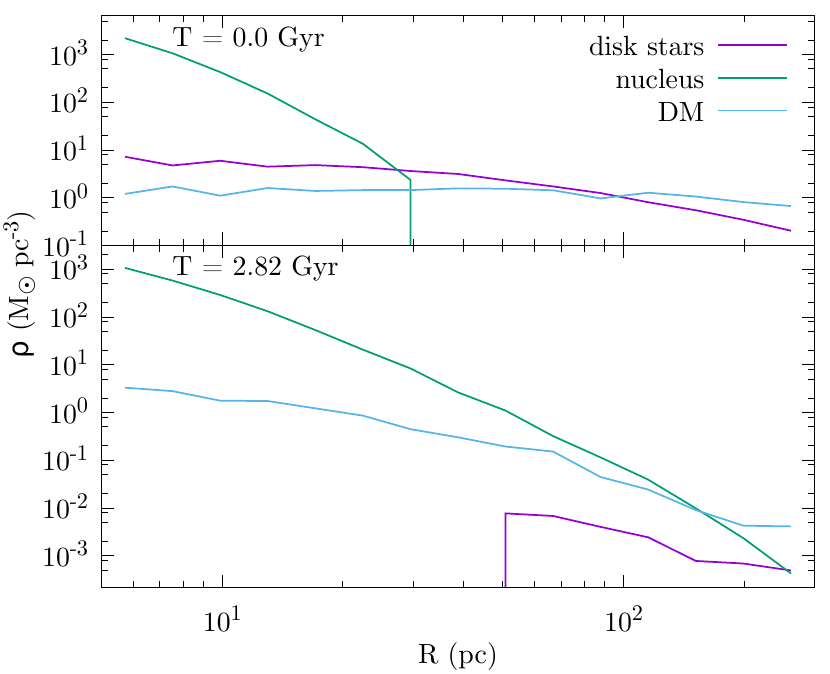}
	\caption{The radial mass density profile of disk stars (purple), nucleus stars (green) and DM (blue) of
	model \csname rundmg118a \endcsname\ at $T = {0.0 ~\rm Gyr}$ (upper panel) and $T = {2.82 ~\rm Gyr}$ (lower panel).
	The sudden cutoff for the nucleus in the upper panel is a binning effect.}
	\label{fig_DensProf}
\end{figure}

\begin{figure}
	\includegraphics{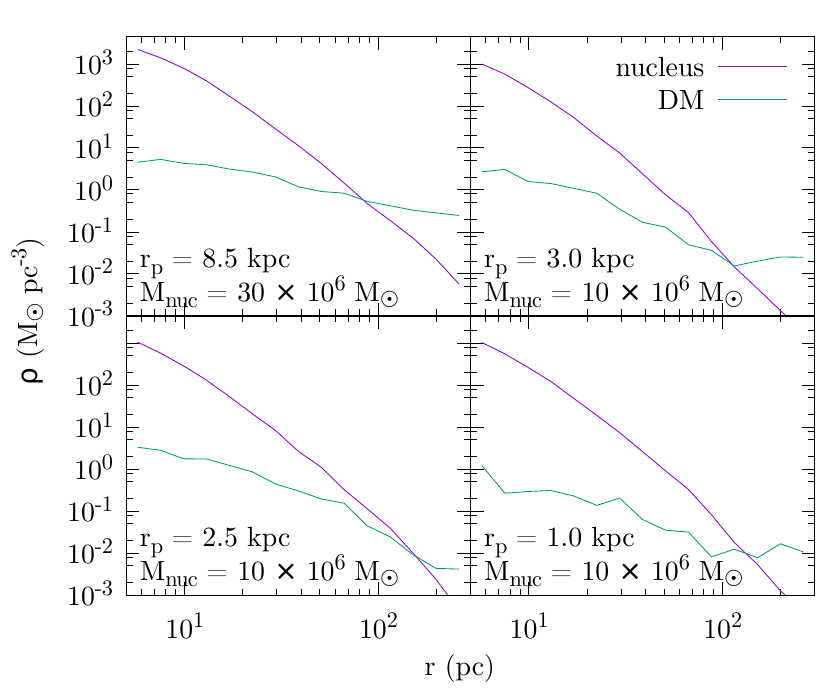}
	\caption{The radial density profiles of different models at $T = {2.82~\rm Gyr}$. From the top left to the bottom right: \csname rundmg119 \endcsname, \csname rundmg120 \endcsname, \csname rundmg123 \endcsname\ and \csname rundmg127 \endcsname.}
	\label{fig_MultiDensProfs}
\end{figure}

\begin{figure}
	\includegraphics{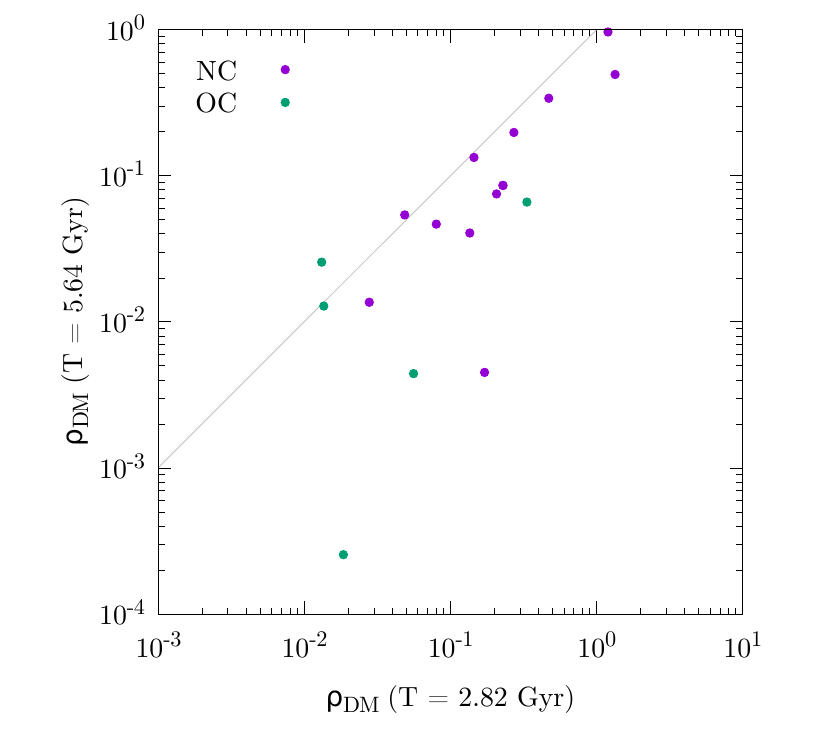}
	\caption{The average DM density (in ${\rm M_\odot pc^{-3}}$) within the inner 100 pc at $T = {5.64~\rm Gyr}$ over the DM density at $T = {2.82~\rm Gyr}$.
		The identity is shown in grey.}
	\label{fig_DensComp}
\end{figure}

\begin{figure}
	\includegraphics[scale=0.34]{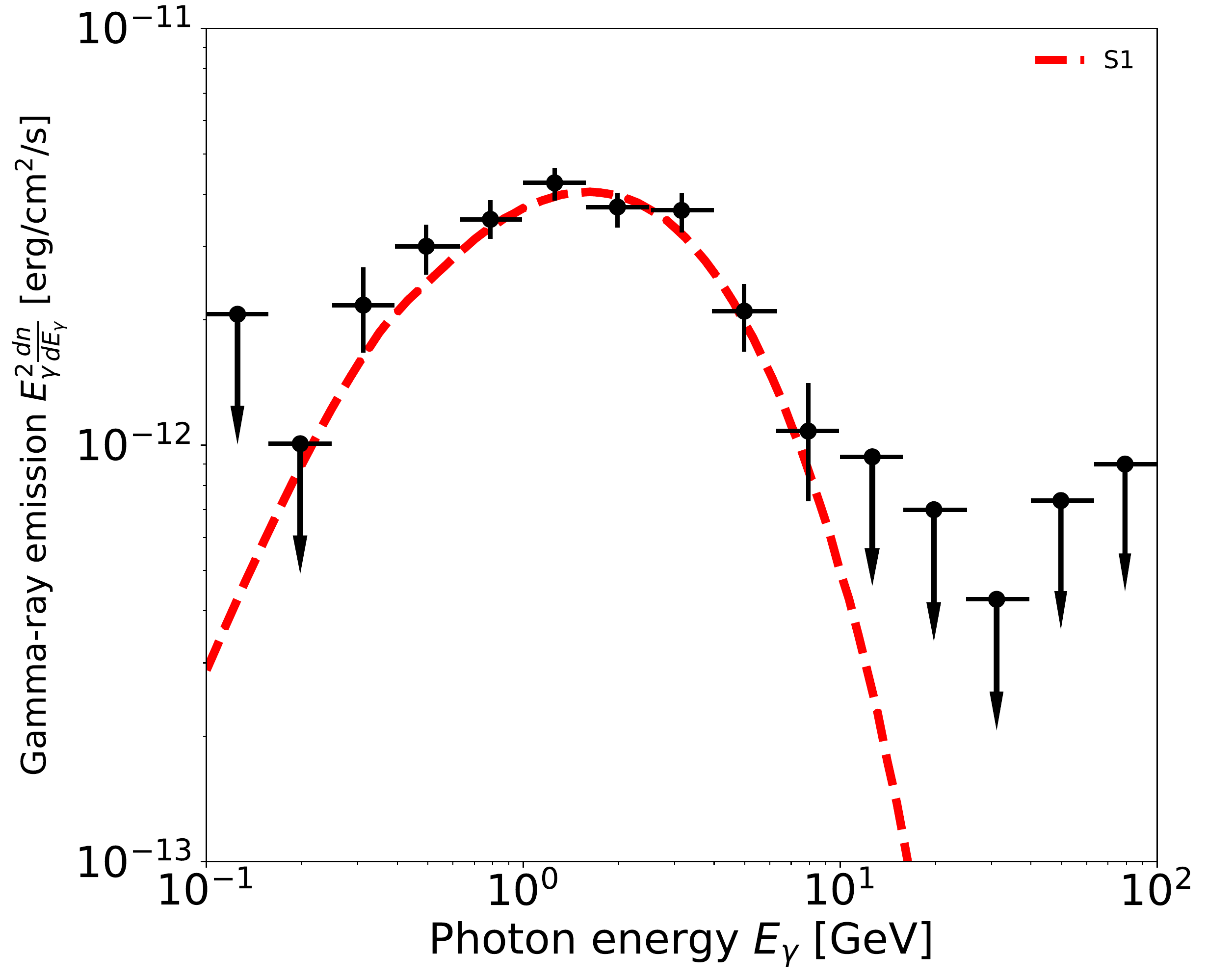}
	\caption{The $\gamma$-ray spectrum of simulated data from model \csname rundmg118a \endcsname\ (red line) and compared to observations of $\omega$ Cen (black dots).
	The observational data is taken from \citet{2019arXiv190708564B} who integrated the data from \citet{2019arXiv190210045T} over 10 years.}
	\label{fig_GammaFlux}
\end{figure}

\section{Results}

Fig. \ref{fig_DensProf} shows how the radial density profiles of \ac{DM}, stellar disk, and nucleus evolve with time during tidal disintegration of the dwarf in the model \csname rundmg118a \endcsname\ with $M_{\rm nuc} = {1 \times 10^7~\rm M_\odot}$ and $R_{\rm peri} = {2~\rm kpc}$.
In this model we saw a strong decrease of $F_{\rm s} (R < 30~\rm pc)$ from $4.92 \times 10^{-2}$ to $0.0$ within the first 2.82 Gyr.
However, if we look at the stellar density of the disk between 40 and 50 pc, we notice that it is only ${\approx 10^{-2}~\rm M_\odot pc^{-3}}$.
With this we would expect only 0.9 stars within the central 40 pc.
Therefore, the sudden cutoff is likely to be caused by the low mass resolution.
Nevertheless, this nucleus is considered stripped, according to our definition.
The \ac{DM} is dynamically relaxed under the presence of the disk in isolation for ${1~\rm Gyr}$ (before the dwarf model is run). The flattened profile seen in the upper plane is due to this dynamical evolution consistent with \citet{2010A&A...514A..47P} and \citet{2015AJ....149..180O}.
Meanwhile, $F_{\rm DM} (R < 30~\rm pc)$ decreased from $1.63 \times 10^{-2}$ to $1.54 \times 10^{-2}$ and to $1.12 \times 10^{-2}$ during the following 2.82 Gyr.
The absolute \ac{DM} density within 30 pc decreases from $2.57$ to $1.66 ~\rm M_\odot pc^{-3}$ and $F_{\rm DM} (R < 100~\rm pc)$ decreases from $0.50$ to $0.06$ during this time.
Only the lighter inner \ac{DM} particles were found within 100 pc after evolution.
Mass segregation can, therefore, not be the cause for the remaining \ac{DM}.

A compilation of density profiles for different models at $T={2.82~\rm Gyr}$ can be seen in Fig. \ref{fig_MultiDensProfs}.
An important observation here is that the \ac{DM} profile steepens again after being exposed to the tidal field of the young \ac{MW}.
It becomes dominant compared to stars close to the \ac{COM} and we also find more \ac{DM} than stars in the inner region at $T={5.64~\rm Gyr}$.
We can see that the central density is lower for lower pericentres, which we will discuss further in the following paragraph.

A compilation of model properties at $T={2.82~\rm Gyr}$ can be seen in Table \ref{table_T20}.
One result is that the \ac{DM} density around the nucleus is smaller for models with a smaller pericentre.
This is due to the tidal forces being stronger closer to the centre of the \ac{MW}.
In \csname rundmg122 \endcsname\ tidal stripping is weaker, because of its large $R_{ini}$.
Theoretically we would expect that heavier nuclei are able to retain more \ac{DM}.
While no such correlation could be found, we cannot exclude it due to our small number of models.
Apart from models \csname rundmg128 \endcsname\ and \csname rundmg133 \endcsname\ the models all show a higher average \ac{DM} density within the inner 30 pc then within a 100 pc radius around the \ac{COM}.
This points to there still being non-stripped \ac{DM} in the nucleus.

The two low-density models \csname rundmg125 \endcsname\ and \csname rundmg127 \endcsname, with smaller NFW c parameter ($=8$), show a significantly lower final \ac{DM} density  ($R < {30~\rm pc}$) than the standard model, with $c=16$ but otherwise similar parameters.
This implies that there is a dependency between the initial and the final \ac{DM} density.
In most of the models with off-centre \acp{GC} disk stars and \ac{DM} are stripped rapidly.
This leads to a \ac{DM} density of less then ${0.2~\rm M_\odot pc^{-3}}$ within the central 30 pc and only a few hundredths of ${\rm M_\odot pc^{-3}}$ within the central 100 pc after 2.82 Gyr.
This result can be understood easiest by viewing the nucleus as being stripped from the galaxy due to its large distance from the dwarf's \ac{COM} and the lack of timefor it to spiral in due to dynamic friction.
In \csname rundmg134 \endcsname, the massive GC can spiral into the central region before the disintegration of its host dwarf, because the pericenter is quiet large and thus tidal stripping is significantly weaker.

Fig. \ref{fig_DensComp} shows a comparison of the \ac{DM} densities at two different times.
Most of the points are below the identity which means that models in general lose \ac{DM} slowly due to tidal stripping during the long term dynamical evolution of the nuclei.
Again we can see that the models with an off centre \ac{GC} instead of a nucleus have on average far less \ac{DM} than the other models.
The exception to this is again \csname rundmg134 \endcsname\ which is visible as the green point at 0.3.

Fig. \ref{fig_GammaFlux} shows the $\gamma$-ray energy spectrum calculated from \ac{DM} annihilation via the $b\bar{b}$ channel in the case of model \csname rundmg118a \endcsname.
In this case, the estimated J-factor value is $J(0.7^{\circ})=1.78\times 10^{22}$”~GeV$^2$~cm$^{-5}$.
Comparing with the observed energy flux of $\omega$ Cen based on Fermi-LAT data \citep[visible as dots in the Fig. \ref{fig_GammaFlux};][]{2019arXiv190708564B,2019arXiv190210045T}, \csname rundmg118a \endcsname\ can explain the observed $\gamma$-ray emissions from \ac{DM} annihilation.
To estimate the size and mass we choose a cutoff density of ${10~\rm M_\odot pc^{-3}}$.
This gives us a radius of  30 pc and a mass of $6.8 \times 10^6~\rm M_\odot pc^{-3}$.
This mass is a little below the highest estimate found in literature for $\omega$ Cen's mass of $7.13 \times 10^6~\rm M_\odot$ \citep{1991ApJ...381..147R}.
However, other sources give significantly lower values i.e. $4.55 \times 10^6~\rm M_\odot$ \citep{2013MNRAS.429.1887D}.
Additionally, the small pericenter distance of \csname rundmg118a \endcsname\ is consistent with corresponding observations.
Although \csname rundmg119 \endcsname\ shows a high central density of \ac{DM} in the \ac{GC}, its $R_{\rm peri}$ is too large for $\omega$ Cen.
This could be a good model for the outer Galactic \acp{GC} with \ac{DM}.
\csname rundmg123 \endcsname, \csname rundmg124 \endcsname, and \csname rundmg126 \endcsname\ also shows high \ac{DM} densities ($\approx 1.5~\rm M_\odot pc^{-3}$ within $30~\rm pc$ and $\approx 0.2~\rm M_\odot pc^{-3}$ within $100~\rm pc$) and small
$R_{\rm peri}$ so that they can be the reasonable model for $\omega$ Cen.

However, not all of the present models
show the required high-density \ac{DM} within the \acp{GC}, because the final \ac{DM} densities
within the central 30 pc depend on the model parameters. For example, \csname rundmg121 \endcsname, which has
a low $M_{\rm nuc}$, shows ${\rho_{\rm dm}}$ of ${0.19 ~\rm M_\odot pc^{-3}}$,  which means that the $\gamma$-ray emission from \ac{DM} annihilation should be too weak owing to
the dependence of the emission flux on the \ac{DM} density squared.
Similarly, the low density models and the models with an off-centre \ac{GC} instead of a nucleus show a very low final \ac{DM} density.
Thus, the large range of the \ac{DM} densities in simulated massive \acp{GC}
suggests that (i) the observed fluxes of gamma ray emission from 47 Tuc and $\omega$ Cen
could be possibly explained by \ac{GC} formation from stripped nuclei but (ii) it is also possible
that the \ac{DM} density in \acp{GC} is not high enough to reproduce
the observed $\gamma$-ray emission if they originate from dwarfs with lower \ac{DM} densities.

\section{Discussion and Conclusion}

We have shown that massive \acp{GC} like $\omega$ Cen can still contain a significant amount of
\ac{DM}, if they originate from nuclei of massive dwarf galaxies.
Also we have shown that \acp{GC} formed well outside the central regions of their host dwarfs
can have no \ac{DM} after they are stripped from the host, even if they are
massive at their birth.
We therefore suggest that the formation sites of \acp{GC} in their hosts rather than
their original masses can determine whether they can contain \ac{DM} thus be
sources of $\gamma$-ray emission from \ac{DM} annihilation.

A number of the Galactic \acp{GC} are observed to have large stellar halos \citep[e.g.][]{2007A&A...466..181C,2009AJ....138.1570O}, and recent numerical simulations
have shown that these stellar halos can be explained, if the \acp{GC} are stripped
nuclei of defunct dwarf galaxies \citep{2012MNRAS.419.2063B}.
These previous studies combined with the present results therefore suggest that there can be other possible candidates of \acp{GC} with \ac{DM}.
On the other hand \citet{2009MNRAS.396.2051B} found no evidence for the presence of substantial \ac{DM} in NGC2419.
How common \ac{DM} is in \ac{GC} remains, therefore, up for debate.
Since these clusters are not so close to us,
the future Cherenkov Telescope Array will be ideal to detect the $\gamma$-ray signals of
\ac{DM} annihilation from these clusters.

Although we have demonstrated that the observed $\gamma$-ray flux in $\omega$ Cen is consistent
with the threshing formation scenario,
it is yet to be determined whether the gamma-ray observation can be explained better by
\ac{DM} annihilation or by milli-second pulsars. One way to distinguish between the
two competing scenario is to observe $\omega$ Cen in radio wavelengths \citep[e.g.][]{2019arXiv190708564B}.
It is thus our future study to investigate the expected radio properties of massive \acp{GC} with
a significant amount of \ac{DM} like $\omega$ Cen based on our dynamical models.

\section*{Acknowledgements}

We would like to thank the anonymous referee for their helpful comments, which greatly improved the paper.
HW wishes to thank ICRAR for their hospitality during her stay there.
This work was supported by JSPS KAKENHI Grant Numbers, 18H04359 \& 18J00277 for KH.

\bibliographystyle{mn2e}
\bibliography{dmg}

\end{document}